\documentclass[twocolumn,aps]{revtex4-1}
\usepackage[latin9]{inputenc}
\usepackage{amsmath}
\usepackage{amssymb}
\usepackage{graphicx}
\usepackage{wasysym}
\usepackage{esint}

\makeatletter

\@ifundefined{textcolor}{}{%
\definecolor{BLACK}{gray}{0}
\definecolor{WHITE}{gray}{1}
\definecolor{RED}{rgb}{1,0,0}
\definecolor{GREEN}{rgb}{0,1,0}
\definecolor{BLUE}{rgb}{0,0,1}
\definecolor{CYAN}{cmyk}{1,0,0,0}
\definecolor{MAGENTA}{cmyk}{0,1,0,0}
\definecolor{YELLOW}{cmyk}{0,0,1,0}
}

\usepackage{upgreek}

\@ifundefined{textcolor}{}{%
\definecolor{BLACK}{gray}{0}
\definecolor{WHITE}{gray}{1}
\definecolor{RED}{rgb}{1,0,0}
\definecolor{GREEN}{rgb}{0,1,0}
\definecolor{BLUE}{rgb}{0,0,1}
\definecolor{CYAN}{cmyk}{1,0,0,0}
\definecolor{MAGENTA}{cmyk}{0,1,0,0}
\definecolor{YELLOW}{cmyk}{0,0,1,0}
}

\usepackage{color}

\def\NOT(#1,#2){\OneQubitGate(#1,#2){$X$}}

\@ifundefined{showcaptionsetup}{}{%
\PassOptionsToPackage{caption=false}{subfig}}

\makeatother

\begin{document}

\title{Magnetocaloric effect as a signature of quantum level-crossing for
a spin-gapped system}

\author{Tanmoy Chakraborty$^{1,2}$, Chiranjib Mitra$^{1}$ }

\affiliation{$^{1}$Indian Institute of Science Education and Research (IISER)
Kolkata, Nadia-741246, West Bengal, India}

\affiliation{$^{2}$Institute for Materials Research (IMO), Hasselt University,
Wetenschapspark 1, B-3590 Diepenbeek, Belgium}
\begin{abstract}
Recent research dealing with magnetocaloric effect (MCE) study of
antiferromagnetic (AFM) low dimensional spin systems have revealed
a number of fascinating ground-state crossover characteristics upon
application of external magnetic field. Herein, through MCE investigation
we have explored field-induced quantum level-crossing characteristics
of one such spin system: $NH_{4}CuPO_{4}\cdotp H_{2}O$ (NCP), an
AFM spin 1/2 dimer. Experimental magnetization and specific heat data
are presented and the data have been employed to evaluate entropy,
magnetic energy and magnetocaloric properties. We witness a sign change
in magnetic Grüneisen parameter across the level-crossing field $B_{C}$.
An adiabatic cooling is observed at low temperature by tracing the
isentropic curves in temperature-magnetic field plane. Energy-level
crossover characteristics in NCP interpreted through MCE analysis
are well consistent with the observations made from magnetization
and specific heat data. 
\end{abstract}
\maketitle
Studying low dimensional quantum antiferromagnets (AFM) has been at
the forefront of both experimental and theoretical condensed matter
physics due to novel nature of their ground states. This has led to
a number of fascinating physical properties \citep{giamarchi2004quantum,lemmens2003magnetic,arnesen2001natural,das2013experimental,giamarchi2008bose,he2017quantum,breunig2017quantum,rahmani2014anyonic}.
Spin-gapped compounds are a class of quantum spin systems, which possess
an energy gap in the excitation spectra \citep{bose2005spin}. Depending
on the nature of the inter-dimer coupling AFM spin dimers exhibit
various kinds of quantum phenomena like Bose-Einstein condensation
of magnons \citep{ruegg2003bose,sebastian2006dimensional}, appearance
of magnetization plateaus \citep{kageyama1999exact,kodama2002magnetic}
etc. under the influence of magnetic field, an external tuning parameter.
Spin dimer systems with significantly weak interdimer interaction
can be considered as independent spin clusters. The energy spectrum
of an AFM spin 1/2 dimer consists of a singlet ground state and a
3-fold degenerate state which upon application of external magnetic
field splits into three states and evolve as the field changes. A
level-crossing between the ground state and the first excited state
occurs when the field is increased through a critical field value.
Thus, the ground state undergoes a qualitative change and the first
excited state becomes the new ground state. Such level-crossing happens
between two pure quantum states which ideally is a zero temperature
phenomena, although it is possible to capture its evidence by measuring
the physical properties of the system at finite temperature.

Here we have demonstrated magnetocaloric effect (MCE) at field-induced
energy level-crossing in $NH_{4}CuPO_{4}\cdotp H_{2}O$ (NCP), a Heisenberg
dimer system. MCE is an intrinsic property of magnetic systems which
in general refers to a change in sample temperature $(T_{Samp})_{S}$
upon adiabatic change of applied magnetic field \citep{tishin2016magnetocaloric}.
Thus MCE is associated with an isothermal change in entropy with field
which is expressed as $(\partial T/\partial H)_{S}=-[T(\partial S/\partial B)_{T}]/C$,
where $C$ is the specific heat in constant field. MCE has been widely
exploited in refrigeration applications both in cryogenic \citep{gschneidnerjr2005recent}
and room temperature \citep{tegus2002transition}. Having larger~change
in entropy $(\Delta S/\Delta B)_{T}$ and $(T_{Samp})_{S}$ in isothermal
and adiabatic process respectively are the characteristics of superior
magnetocaloric materials which have the potential for novel cooling
applications. In this context, paramagnetic salts can act as suitable
substitutes for $^{3}He-^{4}He$ dilution refrigerant in cooling technology.
To avail temperature in mK range \citep{lounasmaaexperimental}, magnetic
refrigeration has potential applications in space technology \citep{chen2014development},
medical sciences \citep{li2012magnetocaloric} and so on.

MCE has been efficiently employed to explore the low temperature behavior
of a variety of systems like rare-earth transition-metal-based magnetic
compounds \citep{das2018giant}, manganite \citep{quintero2010magnetocaloric},
Heusler alloy \citep{synoradzki2018magnetocaloric}, plastic crystals
\citep{Li2019Collosal}, intermetallic compounds \citep{sankar2018magnetocaloric},
low dimensional spins systems like spin chain \citep{zhitomirsky2004magnetocaloric,honecker2009magnetocaloric,breunig2017quantum},
spin ladders \citep{amiri2014thermodynamics} and so on. It has been
observed that an enhancement in MCE occurs being triggered by geometric
frustration in different spin lattice systems which can facilitate
cryogenic cooling applications \citep{zhitomirsky2003enhanced}. For
systems undergoing a level crossing under the influence of magnetic
field, study of MCE can reveal novel qualitative behavior of various
thermodynamic parameters in the vicinity of the quantum critical point.
A few experimental studies have focused on exploring quantum critical
properties by studying MCE of spin 1/2 chain compounds \citep{xiang2017criticality,wolf2011magnetocaloric,breunig2017quantum,sharples2014quantum}.
In the present work, MCE is investigated for NCP by measuring magnetic
and thermal properties at finite temperature. Main motivations of
the present study is exploring the thermodynamic behavior of NCP in
detail and witnessing the field-induced level crossing by capturing
magnetocaloric response. We have also discussed the possibility for
application of NCP as an adiabatic demagnetization refrigerant. 

\begin{figure}
\includegraphics[width=0.8\columnwidth]{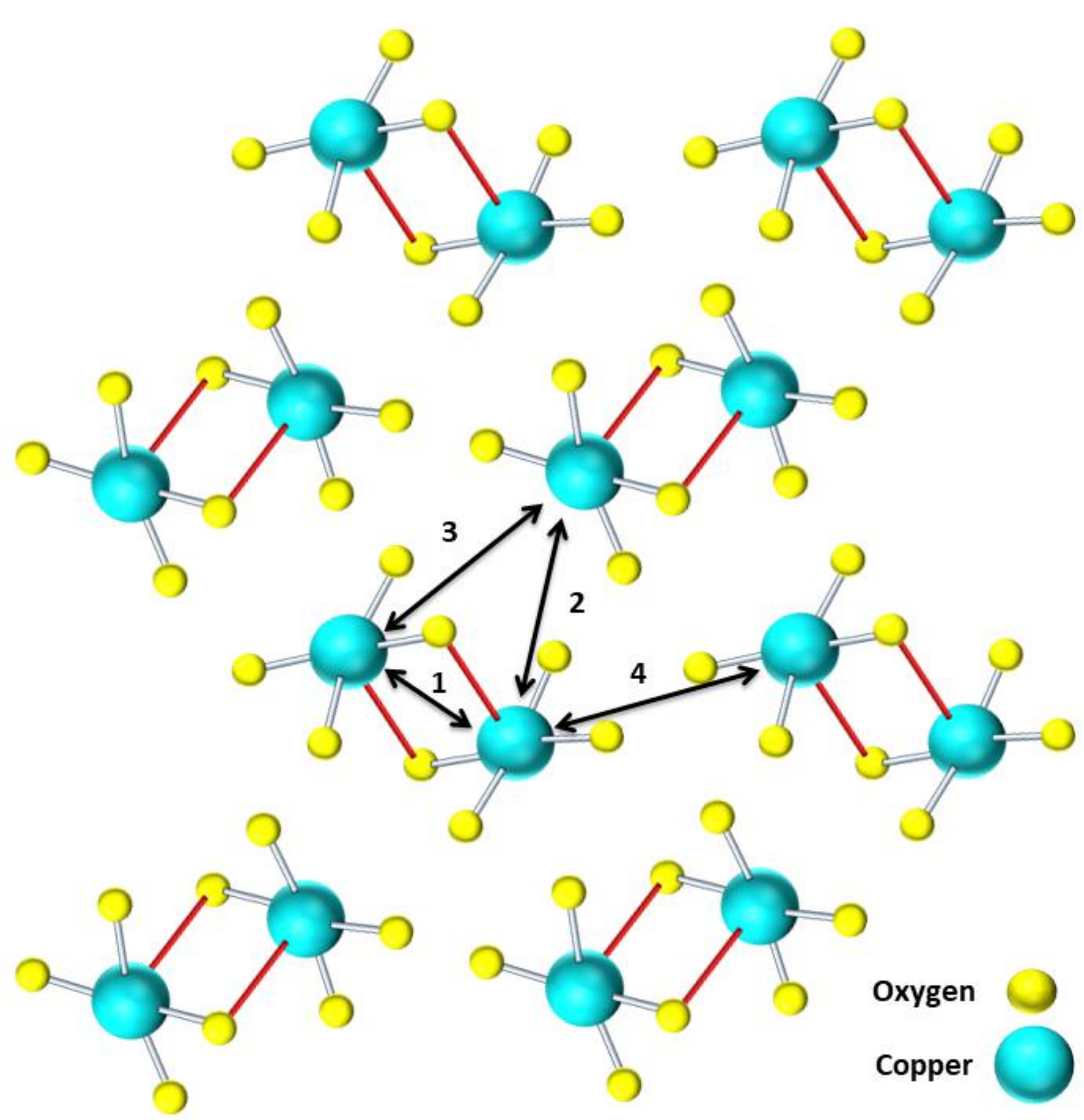}\caption{A schematic diagram of $Cu_{2}O_{8}$ dimers distributed in a layer
in $NH_{4}CuPO_{4}\cdotp H_{2}O$ crystal. Four spin exchange pathways
1, 2, 3 and 4 are shown by the black arrows, the oxygen and the copper
atoms are represented by the yellow and cyan spheres respectively.
The details of the crystal structure can be found in ref. \citep{koo2008correct}.
\label{fig: crystal-structure}}
\end{figure}

The crystallographic analysis of the compound showed layers of centrosymmetric
$Cu_{2}O_{8}$ dimers which are cross connected with each other by
phosphate tetrahadra \citep{pujana1998synthesis}. Fig. \ref{fig: crystal-structure}
schematically exhibits the arrangement of the $Cu_{2}O_{8}$ dimers
in a layered structure and the four spin exchange (EP) pathways. Spin
dimer analysis through extended Hückel tight binding calculations
\citep{koo2008correct} combined with crystallographic analysis \citep{pujana1998synthesis}
suggest, and it has been experimentally proven \citep{chakraborty2014signature}
that being the intradimer interaction (EP1) significantly larger than
the interdimer interactions (EP2, 3 and 4), the system can be best
described by isolated spin 1/2 Heisenberg dimer model, where the two
spins are antiferromagnetically coupled to each other via a strong
intradimer exchange coupling constant $J$. Hence, one can ignore
the terms associated with EP2, 3 and 4 in the Hamiltonian. The Hamiltonian
for a dimer can be expressed as \citep{parkinson2010introduction}:

\begin{equation}
H_{dimer}=J\overrightarrow{S_{1}}.\overrightarrow{S_{2}}+g_{e}\mu_{B}B(S_{1}^{Z}+S_{2}^{Z}),\label{eq:Hamiltonian}
\end{equation}

Here $\overrightarrow{S_{i}}$($i$=1,2) is the vector spin operator
at $i$-th site, $g_{e}$ is the Landé g-factor, $\mu_{B}$ is the
Bohr magneton and $B$ is the applied magnetic field. $J$ is expressed
in unit of $K$. The energy spectrum of the system consists of a singlet
ground state $\frac{1}{\sqrt{2}}(\uparrow\downarrow-\downarrow\uparrow)$
with energy eigenvalue $\frac{-3J}{4}$ and triplet excited states
$\uparrow\uparrow$, $\downarrow\downarrow$ and $\frac{1}{\sqrt{2}}(\uparrow\downarrow+\downarrow\uparrow)$
with eigenvalues $\frac{J}{4}-g_{e}\mu_{B}B$, $\frac{J}{4}+g_{e}\mu_{B}B$
and $\frac{J}{4}$ respectively \citep{parkinson2010introduction}. 

NCP was synthesized and Greenish-blue single crystals were crystallized
following the preparation route described in ref. \citep{chakraborty2014signature}.
To perform a detail magneto-thermal analysis of NCP, magnetization
and specific heat measurements were carried out in the temperature
range, where the short range AFM correlations are persistent, and
in the magnetic field range, where one can capture the closing and
reopening of the spin excitation gap. Temperature and magnetic field
dependent magnetic measurements were performed in a SQUID (Superconducting
Quantum Interference Device) based magnetometer by Quantum Design,
USA in the temperature and magnetic field range of 2-10.2K and 0-9T
respectively. In the same temperature and field range we performed
the heat capacity measurements of NCP by means of AC (alternating
current) calorimetry technique in a cryogen-free magnetic system manufactured
by Cryogenic limited, UK. 

Fig. \ref{fig:M, Cp, U and S data} summarizes the results of the
thermodynamic measurements and shows some of the measured data along
with the theoretical curves for dimer model. Temperature depnedent
magnetization ($M)$ data measured at an applied field of 0.1T is
fitted to the analytical expression for spin 1/2 dimer model described
in ref. \citep{chakraborty2014signature}. Fig. \ref{fig:M, Cp, U and S data}(a)
shows the $M(T)$ data along with the fitted curve. Fig. \ref{fig:M, Cp, U and S data}(b)
depicts the magnetization isotherms measured at 2K, 3K, 6K, 8K and
10.2K and their fit to the expressions for corresponding magnetization
isotherms derived for spin 1/2 dimer Hamiltonian shown in Eq. (\ref{eq:Hamiltonian})
and discussed in ref. \citep{das2013experimental}. The fitting analyses
yielded exchange coupling constant $J$=5K. A step-like behavior associated
with a jump from one plateau to another could be observed in the magnetization
curves especially at lower temperature which corresponds to the level-crossing
of singlet ground state and the excited state $\uparrow\uparrow$.
The peak values in the field derivative of magnetization $(\partial M/\partial H)_{T}$
isotherms in Fig. \ref{fig:M, Cp, U and S data}(c) indicate that
the critical field is $\sim$7T. The jump in $[M(H,T),H]_{T}$ curves
gets less pronounced and the peaks in $(\partial M/\partial H)_{T}$
curves get more broadened at higher temperature as the relative weight
of the singlet state decreases in the statistical mixture of all four
states at higher temperatures \citep{chakraborty2015experimental,chakraborty2014signature}.
Magnetic contribution to the molar specific heat ($C_{m}$) data has
been extracted by subtracting the lattice part from the measured specific
heat data following the similar way described in ref. \citep{chakraborty2014signature}.
We obtained Debye temperature $\theta_{D}$= 101.9K. The measured
specific heat $C(T)$ can be expressed as an addition of three terms:
$C(T)=C_{m}(T)+\gamma T+\beta T^{3}$ where $\gamma$ is Sommerfeld
coefficient and $\beta$ is responsible for lattice contribution.
We showed earlier that NCP being a magnetic insulator, electronic
specific heat does not have any contribution to $C(T)$\citep{chakraborty2014signature}.
When we plotted the experimental data $\frac{C(T)-C_{m}(T)}{T}$ vs.
$T^{2}$, it turned out to be a straight line that passes through
the origin which proves that $\gamma=0$ for NCP. $C_{m}$ vs. $T$
data for fixed magnetic field values of 0T, 2T, 3T, 4T and 5T were
fitted to the expression for dimer model $C_{m}=1/k_{B}T^{2}(\left\langle H_{dimer}^{2}\right\rangle -\left\langle H_{dimer}\right\rangle ^{2})$
\citep{singh2013experimental} for the corresponding field values
with $J$ as the fitting parameter. We obtained $J$=5K which is consistent
with the earlier analysis. The experimental data along with the fitted
curves are shown in Fig. \ref{fig:M, Cp, U and S data}(d). The broad
peaks in $[C_{m}(H,T),T]_{H}$ plots signify Schottky-like anomaly
which arises due to gradual occupation of the energy states upon increasing
temperature \citep{chakraborty2014signature,singh2013experimental}.
An excellent match between the experimental $C_{m}(H,T)$ and $M(H,T)$
data, and their corresponding theoretical fits ascertain that NCP
is a very good representative of Heisenberg spin 1/2 dimer model. 

\begin{figure*}

\begin{centering}
\includegraphics[width=1\paperwidth]{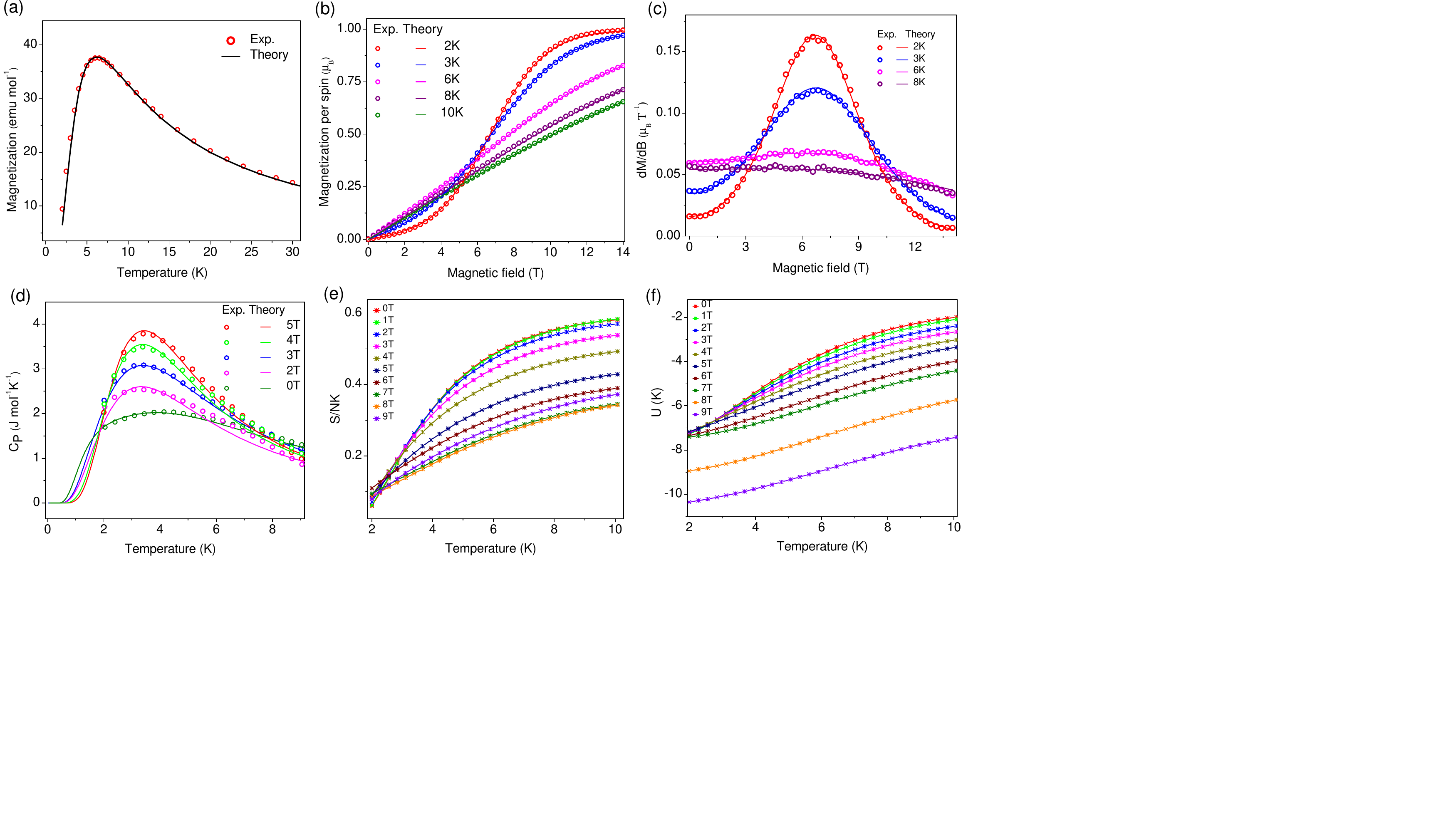}\caption{(a) Temperature dependent magnetization data measured at an applied
magnetic field of 0.1T. (b) Field dependent isothermal magnetization
data at different temperatures, (c) field derivative of the magnetization
isotherms and (d) temperature dependent magnetic specific heat data
at different fields along with the theoretical curves for dimer model
as decribed in the main text. The symbols indicate the data and the
solid lines signify theoretical curves. (e) and (f) show estimated
entropy and internal energy respectively from experimental specific
heat data at various fields as mentioned in the legend.\label{fig:M, Cp, U and S data}}
\par\end{centering}
\end{figure*}

Next, estimation of magneto-thermal properties like magnetic energy
and entropy are performed using the above-presented data, important
magneto-caloric parameters have been calculated and their distinct
features have been discussed. $C_{m}(H,T)$ vs $T$ data-sets measured
at different fields have been substituted in the equation $S_{m}(T)=(S_{m})_{2K}+\varint_{2K}^{10.2K}[C_{m}(H,T')/T']dT'$
\citep{chakraborty2015study} to calculate magnetic entropy $S_{m}(T)$
as a function of temperature for different constant fields. $(S_{m})_{2K}$
for corresponding fields have been determined theoretically for dimer
model and incorporated in the integration. The entropy of a dimer
gradually increases as the system evolves from a populated singlet
state at lower $T$ to the configuration where of the all four eigenstates
are thermally populated at higher $T$ with the occupancy given by
the Boltzmann factor. When $B=J/g_{e}\mu_{B}$, the ground state has
a two fold degeneracy which is associated with an amount of $Rln2$
change in entropy per dimer where $R$ is the universal gas constant.
However, when $B\neq J/g_{e}\mu_{B}$, the entropy reaches a saturation
value of $Rln4$. This is in good analogy with our experimental data
exhibited in Fig. \ref{fig:M, Cp, U and S data}(e). By using the
equation $U_{m}(T)=(U_{m})_{2K}+\varint_{2K}^{10.2K}C_{m}(H,T')dT'$
\citep{chakraborty2015study} numerical integration on $C_{m}(H,T)$
data is carried out to calculate magnetic energy $U_{m}(T)$. Fig.
\ref{fig:M, Cp, U and S data}(f) exhibits $U_{m}(T)$ vs. $T$ for
different magnetic field values. The constant $(U_{m})_{2K}$ for
different magnetic field values are calculated theoretically. When
we consider the evolution of magnetic energy of a dimer as a function
of field, the ground state energy ($E_{g}=-3J/4$) is constant below
level crossing ($B_{C}\sim7T$). However, when $B>B_{C}$, $E_{g}$($=-g_{e}\mu_{B}B+J/4$)
starts varying with field. This change in $E_{g}$is sharp at $0$K,
but gets more gradual at higher temperature due to thermal population
in all four states. Consequently, the constant field $U(T)$ curves
in Fig. \ref{fig:M, Cp, U and S data}(f) starts from nearly equal
values at 2K whereas the starting values increases in the lower side
as $B$ increases beyond 7T.

\begin{figure}
\includegraphics[width=1\columnwidth]{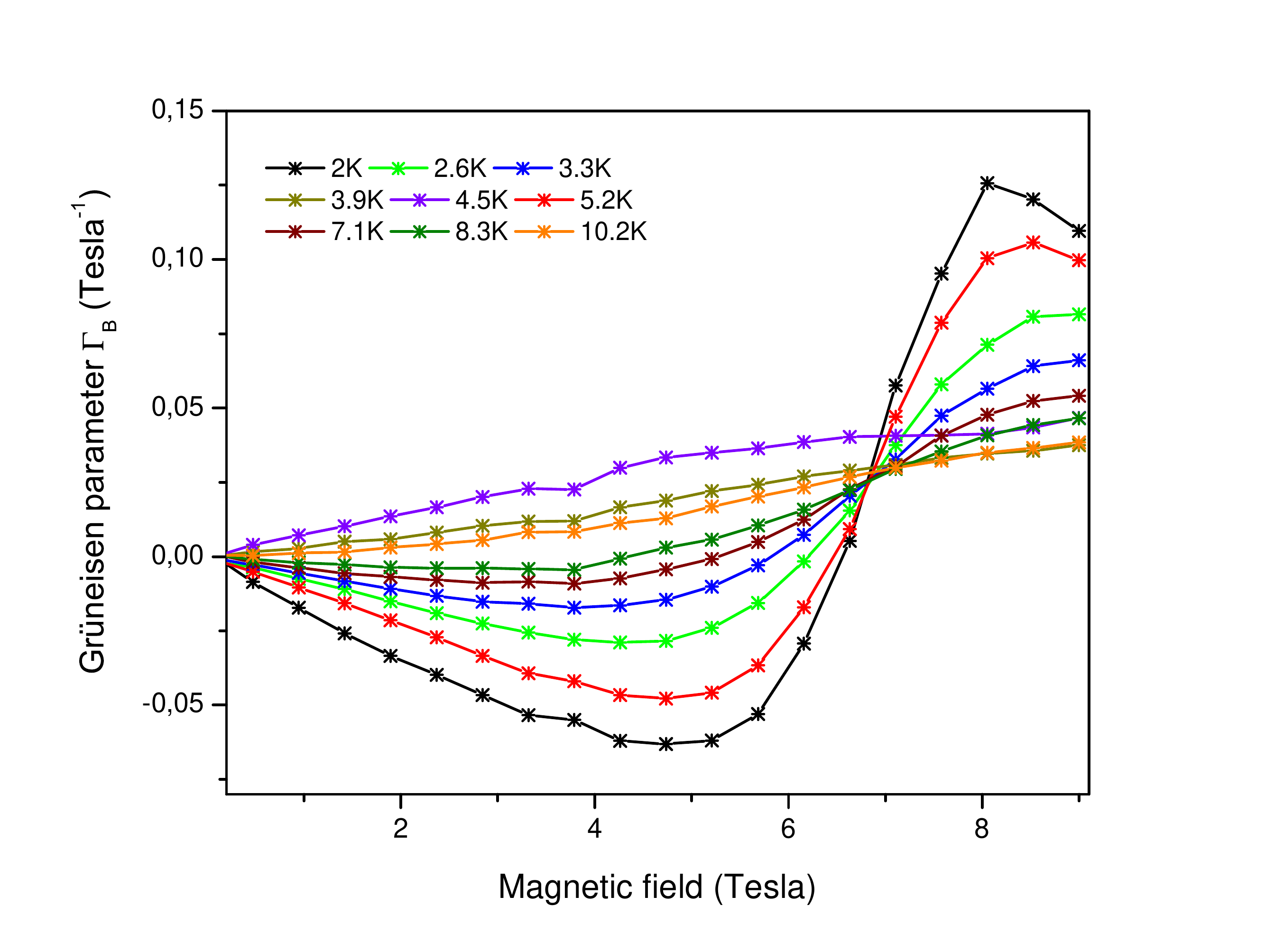}\caption{Experimental magnetic Grüneisen ratio $\Gamma_{B}$ calculated from
measured $C_{m}(H,T)$ and $M(H,T)$ data for NCP. $\Gamma_{B}$ values
are plotted as a function of applied magnetic field for different
constant temperature.\label{fig:Magnetic Grueneisen ratio}}
\end{figure}

An important quantity to capture the crossing of energy levels is
Grüneisen ratio \citep{vovcadlo1994gruneisen,gegenwart2016gruneisen},
that provides a measure of cooling efficiency of the system upon adiabatic
change in magnetic field. In case of magnetic field induced transition,
one can define the magnetic Grüneisen ratio $\Gamma_{B}$ as the adiabatic
change in temperature upon changing the magnetic field, which equals
the ratio of negative temperature derivative of magnetization to heat
capacity at constant field through the following Maxwell's thermodynamic
relations.
\begin{equation}
\Gamma_{B}=(\frac{dM}{dT})_{B}/C_{B}=\frac{1}{T}(\frac{\partial T}{\partial B})_{S}\label{eq:Grueneisen}
\end{equation}
Hence, in fact, an estimation of $\Gamma_{B}$ gives a quantitative
measure of adiabatic MCE. As the system passes through the critical
point upon sweeping the magnetic field, $\Gamma_{B}$ changes its
sign and exhibits divergence. Thus an investigation of adiabatic MCE
can be useful in capturing the critical field. Instead of directly
measuring the adiabatic change in sample temperature with magnetic
field, $\Gamma_{B}$ can alternatively be estimated for a given field
change from $C_{m}(H,T)$ and $M(H,T)$ data by employing the above
Maxwell's relations {[}Eq. (\ref{eq:Grueneisen}){]} \citep{pecharsky2001thermodynamics}.
Using this method we have estimated MCE for NCP. Experimental quantification
of $\Gamma_{B}$ for different temperatures ranging from 2K to 10.2K
were performed and are shown in Fig. \ref{fig:Magnetic Grueneisen ratio}.
As the field increases, the plotted $\Gamma_{B}$ values go through
a minima at $B\sim5T$ and changes its sign from negative to positive
in the vicinity of the level-crossing point. When $\Gamma_{B}$ changes
its sign, an occurrence of maximum thermal entropy happens because
of competing ground states $\frac{1}{\sqrt{2}}(\uparrow\downarrow-\downarrow\uparrow)$
and $\uparrow\uparrow$ in the field region $B<B_{C}$ and $B>B_{C}$
respectively, which consequently creates a frustration in the system.
Upon increasing the field further, $\Gamma_{B}$ passes through a
maximum when $B\sim8T$. Below $B_{C}$, the negative values of $\Gamma_{B}$
signifies adiabatic cooling due to magnetization, whereas an enhancement
in $\Gamma_{B}$ above $B_{C}$ signifies cooling through demagnetization.
The contrast of the $\Gamma_{B}(B)$ curve is higher at lower $T$
because, when $T$ decreases the statistical weight of the singlet
ground state $\frac{1}{\sqrt{2}}(\uparrow\downarrow-\downarrow\uparrow)$
increases in the thermal mixture of the all four states.With decreasing
$T$ the minima and maxima at $B\sim5T$ and $8T$ gets sharper and
eventually would diverge at $T=0$ due to the field induced critical
behavior of Heisenberg spin system NCP. This diverging property of
$\Gamma_{B}$ for NCP in the both sides of the critical point is consistent
with the observation made for $KCuF_{3}$, another spin 1/2 quantum
critical system \citep{wolf2011magnetocaloric}. The sign change in
$\Gamma_{B}$ is more drastic at lower temperatures and centered close
to $B_{C}$ which enables one to identify the level crossing point.
Thus one can note the critical field $B_{C}\sim7$T for NCP. 

\begin{figure}
\includegraphics[width=1\columnwidth]{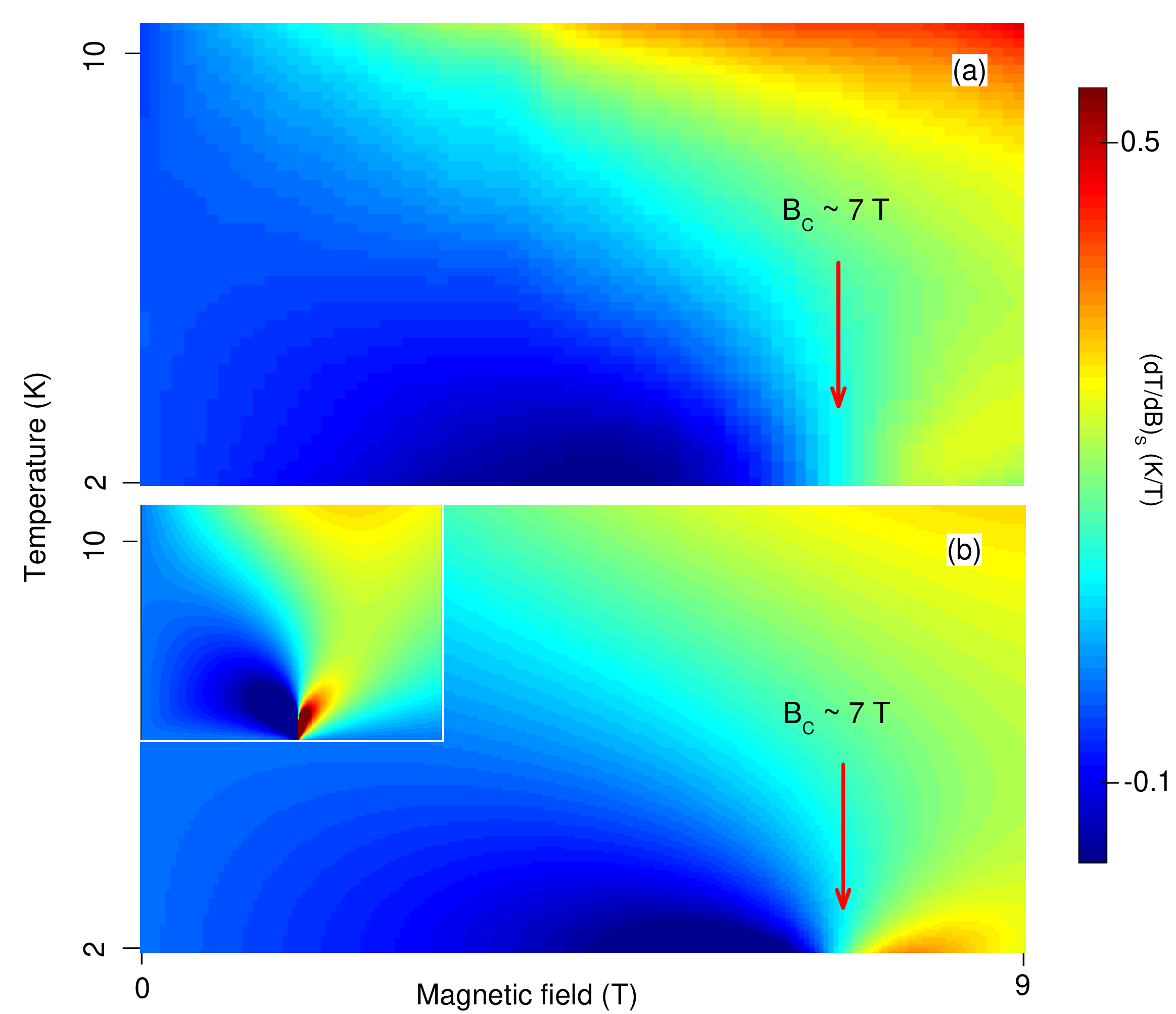}\caption{(a) Experimental and (b) theoretical (dimer) $(dT/dB)_{S}$ represented
in contour plots in $T-B$ plane. Both the plots are shown in same
range: $T$=2K to 10.2K and $B$=0T to 9T. The plot shown in inset
exhibits the features of $(dT/dB)_{S}$ plot for dimer model down
to 0K. \label{fig:dT/dB}}
\end{figure}

An interactive way to trace the sample temperature upon sweeping the
field and thus capturing the critical point is to represent the quantity
$(dT/dB)_{s}$ in a contour plot. We have used the experimental thermodynamic
data in Eq. (\ref{eq:Grueneisen}) to quantify $(dT/dB)_{s}$ and
generated a contour plot shown in Fig. \ref{fig:dT/dB}(a) along with
the simulation for spin dimer in Fig. \ref{fig:dT/dB}(b). The Hamiltonian
mentioned in Eq. (\ref{eq:Hamiltonian}) was used to simulate magnetization
and specific heat, which were eventually used to calculate $(dT/dB)_{s}$
using the Eq. (\ref{eq:Grueneisen}). We used MATLAB for the simulations.
Both the plots are depicted in the same temperature (2 to 10.2K) and
field (0 to 9T) range and they exhibit a very good resemblance. Inset
of Fig. \ref{fig:dT/dB}(b) shows the simulation of $(dT/dB)_{s}$through
$B_{C}$ down to 0K. A minimum in $(dT/dB)_{s}$suggests an enhanced
MCE close to the $B_{C}$ which are evident both from the experimental
and theory plots. 

\begin{figure}
\includegraphics[width=1\columnwidth]{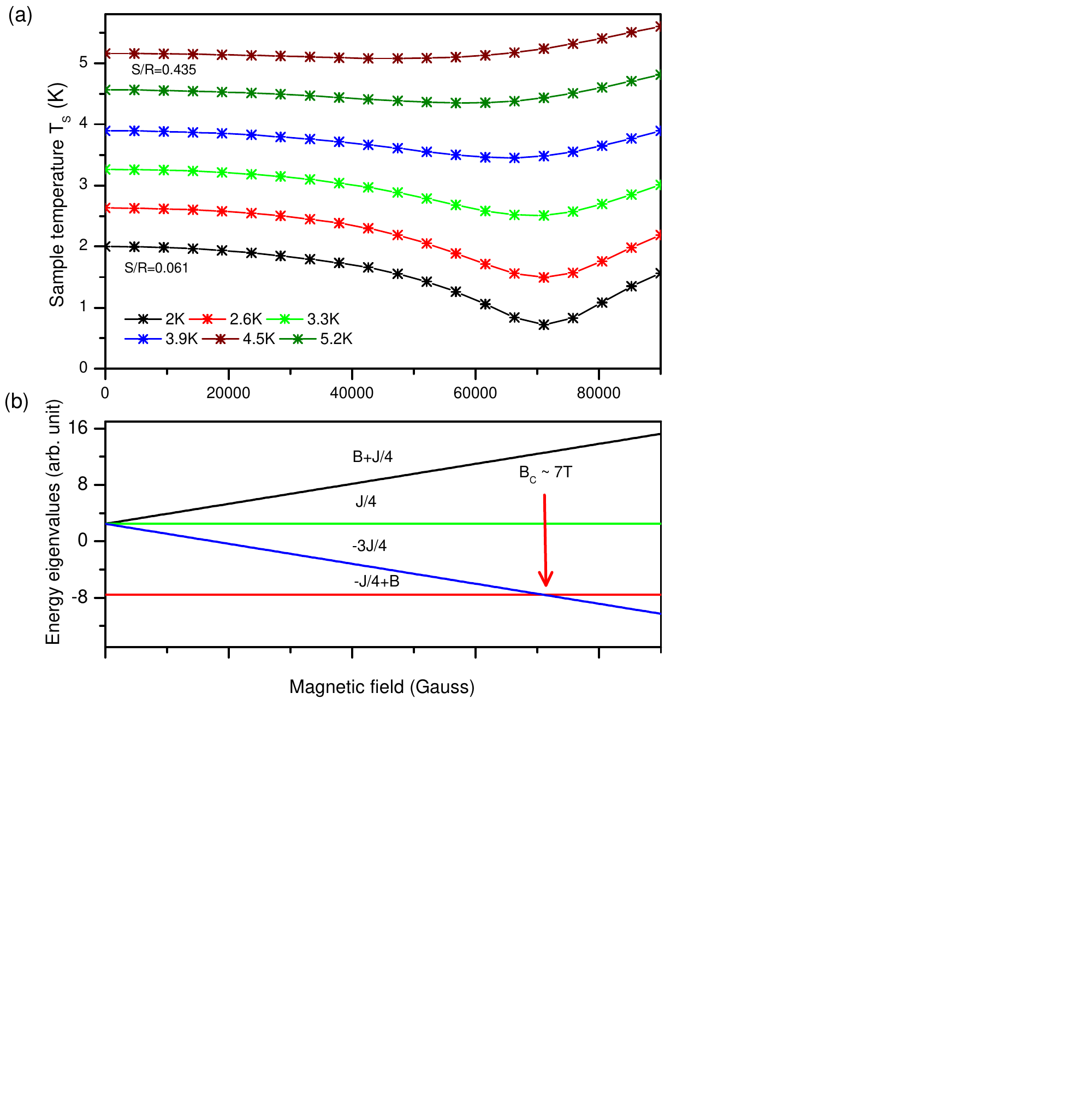}\caption{(a) Experimental isentropes for NCP where $S/R$ varies from 0.061
to 0.435. The curves represent adiabatic cooling of the system upon
sweeping the magnetic field. (b) Simulated energy spectrum for dimer
shows the magnetic field induced energy level crossing around a critical
field value of 7T. At the same field we observed maximum cooling in
the isentrope with lowest initial starting temperature of 2K \label{fig:Cooling}.}
\end{figure}

The observation of substantial variation in $\Gamma_{B}$ with field
from the last analysis indicates that NCP can exhibit significant
change in temperature with small variation of magnetic field which
motivates us to examine its performance as magnetic refrigerant. Using
the experimental thermodynamic data $C_{m}(H,T)$ and $M(H,T)$ we
have quantified the adiabatic variation of sample temperature in response
to change in field by employing the following thermodynamic relation
\citep{pecharsky2001thermodynamics}

\begin{equation}
\bigtriangleup T=T_{f}-T_{i}=-\intop_{H_{i}}^{H_{f}}\frac{T(H)}{C_{m}(H,T)}[\frac{\partial M(H,T)}{\partial T}]_{H}dH\label{eq:dT}
\end{equation}
Here $H_{i}$ and $H_{f}$ are the initial and final magnetic fields
which correspond to initial and final sample temperature $T_{i}$
and $T_{f}$ \citep{tishin1999magnetocaloric}. Quantified experimental
constant-entropy curves for entropy varying from $S/R=0.061$ to $0.435$
correspond to different initial temperatures starting from 2K to 5.2K
are plotted in $T-B$ plane and are shown in Fig. \ref{fig:Cooling}(a).
The plot reveals that the highest variation in $T_{S}$ with field
happens for the curve with $S/R=0.061$ which signifies that cooling
is maximum when the initial value of $T_{S}$ is minimum. Fig. \ref{fig:Cooling}(b)
exhibits the evolution of the four eigenstates with field and the
crossover between the $\frac{1}{\sqrt{2}}(\uparrow\downarrow-\downarrow\uparrow)$
and $\uparrow\uparrow$ states. The lowest entropy isentrope shows
a minimum in the vicinity of the level crossing field signifying a
noticeable magnetocaloric response at the closing of the spin gap
which captures the signature of quantum criticality in NCP. Energy
level crossover properies in NCP interpreted from the MCE is consistent
with the observation made from magnetization data discussed above.
NCP exhibits adiabatic cooling over a wide range of temperature although
the minima in the isentropes get less sharper and the cooling is less
along the curves with higher values of $S/R$ as shown in Fig. \ref{fig:Cooling}(a).
The curve with $S/R=0.435$ shows minimum variation in $T_{S}$ with
field. This MCE study shows that we could achieve a cooling down to
a temperature as low as 0.7K for the minimum initial temperature of
2K.

Although in this report the MCE investigation for NCP is not performed
at sufficiently low temperatures, a notable adiabatic cooling response
for $T_{i}=$ 2K indicates that it can be a possible candidate for
coolant applications in low temperature. By the virtue of being an
AFM critical spin system, it readily has advantages over paramagnetic
salts as coolant: a distinct excitation spectrum of such spin systems
allows low energy magnetic excitations above the level-crossing point
which makes it possible to achieve significant cooling \citep{wolf2011magnetocaloric}.
Specific heat properties of NCP enables the system to enhance the
ability to absorb heat with warming them up in a slow rate. Furthermore,
the system can be prepared following a simple synthesis route and
can be crystallized into regular shaped, mechanically rigid, vapor
and air-stable single crystals in less than one month. These properties
indicate that close to the level-crossing point, NCP can be a possible
material for cooling applications over a wide temperature range. However,
still a few more investigations like studying thermal conductivity
and estimating the achievable base temperature are necessary to test
the efficiency of NCP as a magnetic coolant. We plan to perform these
investigations in details in future.

To summarize, through MCE study certain characteristics of magnetic
field induced level-crossing in an AFM spin gapped system have been
captured. We have carried out a detail study of experimental thermodynamic
properties and showed their analogy to the simulation for Heisenberg
spin 1/2 dimer model. Experimental magnetization, specific heat, magnetic
energy and entropy as a function of temperature and magnetic field
have been investigated. The system exhibits a notable MCE in the vicinity
of the critical magnetic field where the transition between singlet
ground state and triplet first excited state happens. Thus, tracing
the experimental isentropes we could determine the critical field
for NCP which turned out to be $\sim7$T. A noticeable amount of MCE
is observed over a wide temperature range. The demonstration of adiabatic
cooling of NCP opens up a possibility for applying the material as
a refrigerant in coolant technology. 
\begin{acknowledgments}
The authors would like to thank the Ministry of Human Resource Development
(MHRD), Government of India, for funding. 
\end{acknowledgments}

\bibliographystyle{apsrev}
\bibliography{draft}

\end{document}